\documentclass[12pt]{iopart}
\usepackage{graphicx}

\begin{document}

\title[Visibility]{New intensity and visibility aspects of a double loop neutron interferometer}

\author{M. Suda$^{+*}$, H. Rauch$^{*}$ and M. Peev$^{+}$}
\address{$^{+}$ARC Seibersdorf research Ltd., A-2444 Seibersdorf \\
$^{*}$Atomic Institute of the Austrian Universities, Stadionallee 2, A-1020 Wien, Austria\\
e-mail: martin.suda@arcs.ac.at \\ }

\begin{abstract}
Various phase shifters and absorbers can be put into the arms of a
double loop neutron interferometer. The mean intensity levels of
the forward and diffracted beams behind an empty four plate
interferometer of this type have been calculated. It is shown that
the intensities in the forward and diffracted direction can be
made equal using certain absorbers. In this case the
interferometer can be regarded as a $50/50$ beam splitter.
Furthermore the visibilities of single and double loop
interferometers are compared to each other by varying the
transmission in the first loop using different absorbers. It can
be shown that the visibility becomes exactly $1$ using a phase
shifter in the second loop. In this case the phase shifter in the
second loop must be strongly correlated to the transmission
coefficient of the absorber in the first loop. Using such a device
homodyne-like measurements of very weak signals should become
possible.
\end{abstract}


%
\maketitle

\section{Introduction}
Neutron interferometry is already a well known method for
investigation of coherent matter wave properties. Many features of
neutron waves have been analyzed such as coherence and post -
selection effects, the spinor symmetry, the magnetic Josephson
effect, the multi - photon exchange interaction, spin
superposition problems, the Aharonov - Bohm effect, topological
phases, gravitationally induced quantum interference, the Sagnac
effect, the neutron Fizeau effect et cetera [1].
\\
\\
In this paper we concentrate on coherence effects in a double loop
interferometer [2],[3]. Fig.(1) describes an extension of a
standard three - plate neutron interferometer where a second loop
is added by using a second mirror crystal M2 and where various
phase shifters $\Delta$ and absorbers $\alpha$ can be inserted in
each loop. Both loops are coupled via beam $(d)$ and therefore
most attention will be given to the action of an absorbing phase
shift in this beam. We focus on two phase shifters $\Delta_{d}$
and $\Delta_{f}$ and two absorbers $\alpha_{d}$ and $\alpha_{f}$
in beam $(d)$ and $(f)$ respectively because phase shifters or
absorbers in beam $(b)$ are of no additional significance.
\\
\\
The paper is organized as follows. In section $2$ intensity
aspects of the double loop interferometer are considered. In spite
of the non symmetric diffraction properties of neutrons in single
crystals it can be shown that a $50/50$ intensity split-up can be
reached using an absorber in loop $B$. One of the main findings of
the analysis of the double loop interferometer in section $3$ is
that the visibility of $K_{0}$ in the forward beam can be made
very high (ideally $1$), even for very weak input signals. In
section $4$ the results are discussed in detail taking into
account real experimental situations.
\\
\\
The main aim is to find an interaction where small signals
transmitted through phase shifter $\Delta_{d}$ can be detected
with high precision. In this respect it is a search for
homodyne-like detection of weak neutron signals by the constrain
that a symmetric beam splitter does not exist in case of
diffraction from a crystal. The new system consists of a coupled
double loop perfect crystal system and provides a symmetric beam
splitting or can serve as a basic of homodyne neutron detection in
a similar sense as known for photon beams [4].
\begin{figure}
\begin{center}
\includegraphics[angle=-90,width=13cm]{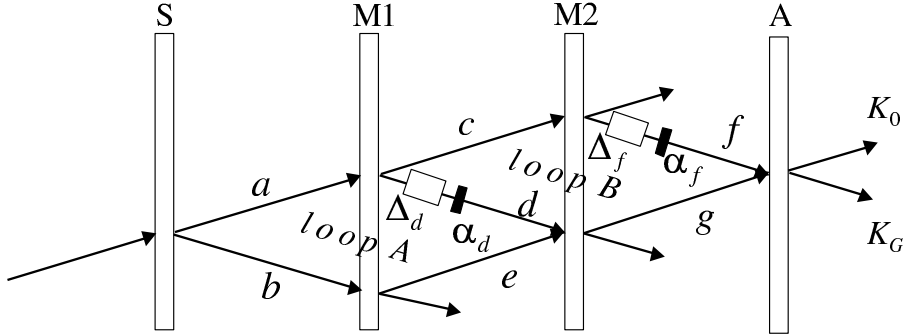}
\end{center}
\caption{\label{label}Double loop neutron interferometer: there
are three beam pathes: $(acf)$, $(adg)$ and $(beg)$. $\Delta_{d}$
and $\Delta_{f}$ are two phase shifters, $\alpha_{d}$ and
$\alpha_{f}$ are two absorbers, $K_{0}$ and $K_{G}$ denote the
intensity in the forward and diffracted direction respectively. S
is the beam splitter crystal, M1 and M2 are two mirror crystals
and A is the analyzer crystal. All four crystal plates have the
same thickness $D$ and have equal distance from each other. The
first loop (loop $A$) is formed by the beams $(adeb)$, the second
loop (loop $B$) by the beams $(cfgd)$. Beam $(g)$ is a
superposition of beam $(ad)$ and beam $(be)$.}
\end{figure}

\section{Intensity aspects}

From the dynamical theory of diffraction the wave function behind
the analyzer crystal in the forward direction (index $0$) which is
a superposition of three waves due to the three pathways (s.
Fig.(1)) reads as ([1],[5])
\begin{equation}
\psi_{0}^{(acf)+(adg)+(beg)}=\psi_{e}e^{i\gamma}v_{0}^{2}v_{G}v_{0}'[e^{ik\Delta_{f}-\alpha_{f}}+e^{ik\Delta_{d}-\alpha_{d}}+1]\,\,\,.
\end{equation}
In this expression $\psi_{e}=u_{0}e^{i\bf{kr}}$ is the incoming
plane wave function with amplitude $u_{0}$, wave vector $\bf{k}$
and position vector $\bf{r}$ impinging on the beam splitter S.
$e^{i\gamma}$ is a phase factor which is of no relevance in the
following. The crystal functions $v_{0}$, $v_{G}$ and $v_{0}'$ are
given as [5]
\begin{equation}
v_{0}=e^{iPD}[cos(\bar{A}\sqrt{1+y^{2}})+iy\frac{sin(\bar{A}\sqrt{1+y^{2}})}{\sqrt{1+y^{2}}}]\,\,\,.
\end{equation}
\begin{equation}
v_{G}=-ie^{iPD}\frac{sin(\bar{A}\sqrt{1+y^{2}})}{\sqrt{1+y^{2}}}\,\,\,,\,\,\,v_{0}'=v_{G}(-y)\,\,\,.
\end{equation}
The index $G$ means the diffracted direction.
$P=-{\pi}(1+y)/\Delta_{0}$ and $\bar{A}={\pi}D/\Delta_{0}$ where
$\Delta_{0}$ is the Pendelloesung length which is
$\Delta_{0}={\lambda}\,cos(\gamma)/|V({\bf{G}})/E|$. $
\lambda=2\pi/k\,$ is the wave length and $\gamma$ is the angle
between the beam direction and the vector perpendicular to the
crystal surface. $|V({\bf{G}})/E|{\approx}10^{-6}$ is the ratio
between the crystal potential $V({\bf{G}})$ and the energy $E$ of
the particles where $\bf{G}$ is the reciprocal lattice vector
corresponding to the reflection plane under consideration.
$\Delta_{0}=0.00641\,cm$ for the $(2,2,0)-$ reflection of $2\AA$
neutrons in a silicon single crystal. The quantity $\bar{A}$ is
proportional to the crystal thickness $D$. The quantity $y$
depends on the deviation from the Bragg angle. For symmetric
diffraction $y$ is defined as
$y=k{\,}sin(2\theta_{B}){\,}(\theta_{B}-\theta)/|V({\bf{G}})/E|$.
$y=0$ gives the exact direction of the Bragg angle
$\theta=\theta_{B}$. The crystal function $v_{0}$ describes the
diffraction property of the wave function through a single crystal
of thickness $D$. $v_{G}$ characterizes the case of diffraction
due to the dynamical theory. The three terms in Eq.(1) form a
superposition of three wave functions which belong to the three
pathways $(acf)$, $(adg)$ and $(beg)$. They have different phase
shifts and absorptions. The crystal functions defined above are
equal for each of the three wave functions, because each beam is
transmitted (function $v_{0}^{2}$) and diffracted (functions
$v_{G}$ and $v_{G}(-y)$) twice respectively.
\\
\\
The wave function in the diffracted direction behind the analyzer
crystal reads
\begin{equation}
\psi_{G}^{(acf)+(adg)+(beg)}=\psi_{e'}e^{i\delta}v_{0}v_{G}[v_{0}v_{G}'e^{ik\Delta_{f}-\alpha_{f}}+v_{G}v_{0}'(e^{ik\Delta_{d}-\alpha_{d}}+1)]\,\,\,.
\end{equation}
Here, $\psi_{e'}=u_{0}e^{i{\bf{k}}_{G}{\bf{r}}}$ is a plane wave
function with wave vector $ {\bf{k}}_{G} $ pointing to the
diffracted direction. $e^{i\delta}$ is a phase factor. The crystal
function $v_{G}'$ is equal to $v_{0}(-y)$. From Eq.(4) it can be
recognized that the three wave functions which are superimposed in
the forward direction (0) have the same but those in the
diffracted direction $G$ do not have the same number of
reflections and transmissions. Therefore an unsymmetrical
intensity behavior can be expected in the diffracted direction, as
shown below.
\\
\\
In order to compute the intensities $K_{0}$ and $K_{G}$ behind the
analyzer crystal the squared moduli of the wave functions, i.e.
$|\psi_{0}^{(acf)+(adg)+(beg)}|^{2}$ and
$|\psi_{G}^{(acf)+(adg)+(beg)}|^{2}$, have to be taken into
account. Because of the thickness of about $0.5$ cm of the crystal
plates in Fig.(1) the quantity $\bar{A}={\pi}D/\Delta_{0}>>1$
cause very rapid oscillations of the terms $sin^{2n}(x)$ which
appear in these expressions. Therefore the mean values of these
trigonometric functions can be used. One gets:
$\overline{sin^{2}(x)}=1/2$, $\overline{sin^{4}(x)}=3/8$,
$\overline{sin^{6}(x)}=5/16$ and $\overline{sin^{8}(x)}=35/128$
[2]. Moreover integration over the variable $y$ must be performed
because of the beam divergence which has to be taken into account.
Finally a normalized Gaussian spectral distribution
\begin{equation}
g(k)=\frac{1}{\sqrt{2\pi}(\delta{k})}exp[\frac{(k-k_{0})^{2}}{2(\delta{k})^{2}}]
\end{equation}
of incoming wave numbers $k$ is assumed. Here $k_{0}$ is the mean
value and $(\delta{k})^{2}$ is the mean square deviation of wave
numbers. The intensities can now be calculated (setting $u_{0}=1$)
as follows (the limits of integration are always $-\infty$ and
$+\infty$):
\begin{eqnarray}
K_{0}=\int{g(k)}[\int{\overline{|\psi_{0}^{(acf)+(adg)+(beg)}|^{2}}}dy]dk=\frac{79\pi}{2048}\{1+e^{-2\alpha_{d}}+e^{-2\alpha_{f}}+\nonumber\\
+2e^{-\alpha_{d}}e^{-(\delta{k})^{2}(\Delta_{d})^{2}/2}cos(\Delta_{d}k_{0})+2e^{-\alpha_{f}}e^{-(\delta{k})^{2}(\Delta_{f})^{2}/2}cos(\Delta_{f}k_{0})+\nonumber\\
+2e^{-(\alpha_{d}+\alpha_{f})}e^{-(\delta{k})^{2}(\Delta_{d}-\Delta_{f})^{2}/2}cos[(\Delta_{d}-\Delta_{f})k_{0}]\}\,\,\,,
\end{eqnarray}
\begin{eqnarray}
K_{G}=\int{g(k)}[\int{\overline{|\psi_{G}^{(acf)+(adg)+(beg)}|^{2}}}dy]dk=\frac{\pi}{2048}\{65(1+e^{-2\alpha_{d}})+\nonumber\\
+417e^{-2\alpha_{f}}+130e^{-\alpha_{d}}e^{-(\delta{k})^{2}(\Delta_{d})^{2}/2}cos(\Delta_{d}k_{0})-\nonumber\\
-158e^{-\alpha_{f}}e^{-(\delta{k})^{2}(\Delta_{f})^{2}/2}cos(\Delta_{f}k_{0})-\nonumber\\
-158e^{-(\alpha_{d}+\alpha_{f})}e^{-(\delta{k})^{2}(\Delta_{d}-\Delta_{f})^{2}/2}cos[(\Delta_{d}-\Delta_{f})k_{0}]\}\,\,\,.
\end{eqnarray}
\begin{figure}
\begin{center}
\includegraphics[angle=0,width=10cm]{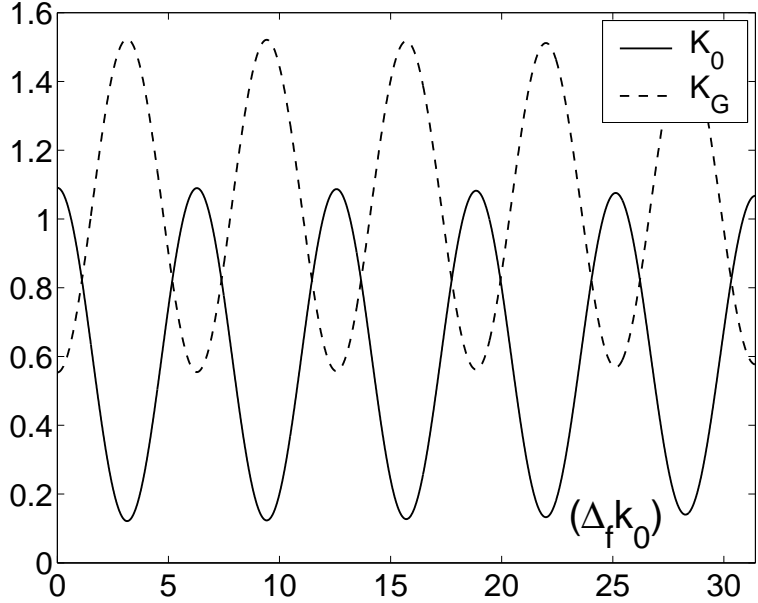}
\end{center}
\caption{\label{label}Intensities $K_{0}$ and $K_{G}$ of Eqs.(6)
and (7) behind the double loop interferometer (s. Fig.(1)) for
$\alpha_{d}=\alpha_{f}=0$, $\Delta_{d}=0$ and
$({\delta}k)/k_{0}=1/100$. The level of oscillations of $K_{0}$ is
below that of $K_{G}$.}
\end{figure}
\\
Fig.(2) plots $K_{0}$ and $K_{G}$ for $\Delta_{d}=0$ and
$\alpha_{d}=\alpha_{f}=0$ as a function of $\Delta_{f}$. If
$\Delta_{f}=0$, $K_{0}=711{\pi}/2048$ and $K_{G}=361{\pi}/2048$.
These results can be found elsewhere [2]. One can recognize that
the mean intensities of the two beams are - in general - different
(just like in single loop interferometry). The question arises if
these intensity levels can be made equal by using special
parameter values. The answer is yes. If one takes the mean values
of $K_{0}$ and $K_{G}$ to be equal, i.e.
$\overline{[K_{0}-K_{G}]}_{\Delta_{f}}=0$, one gets a condition
for $\alpha_{f}$:
\begin{equation}
\alpha_{f}=-\frac{1}{2}ln{\{}\frac{7}{169}[\,1+e^{-2\alpha_{d}}+2\,e^{-\alpha_{d}}e^{-({\delta}k)^{2}({\Delta}_{d})^{2}}cos(\Delta_{d}k_{0})]{\}}{\ge}0\,\,\,\,.
\end{equation}
Any absorption $\alpha$ has to be ${\ge}0$, because the
transmission $T=e^{-2\alpha}$ of a beam obeys the relation
$0{\le}T{\le}1$. In case of $\alpha_{d}=0$ and $\Delta_{d}=0$, the
absorption $\alpha_{f}$ becomes $-(1/2)ln(28/169)=0.8988$. Fig.(3)
demonstrates this example. In this case the double loop
interferometer may be regarded as a $50/50$ beam splitter by using
an absorption element in a suitable arm of the device. This is not
possible using a single loop interferometer.
\begin{figure}
\begin{center}
\includegraphics[angle=0,width=10cm]{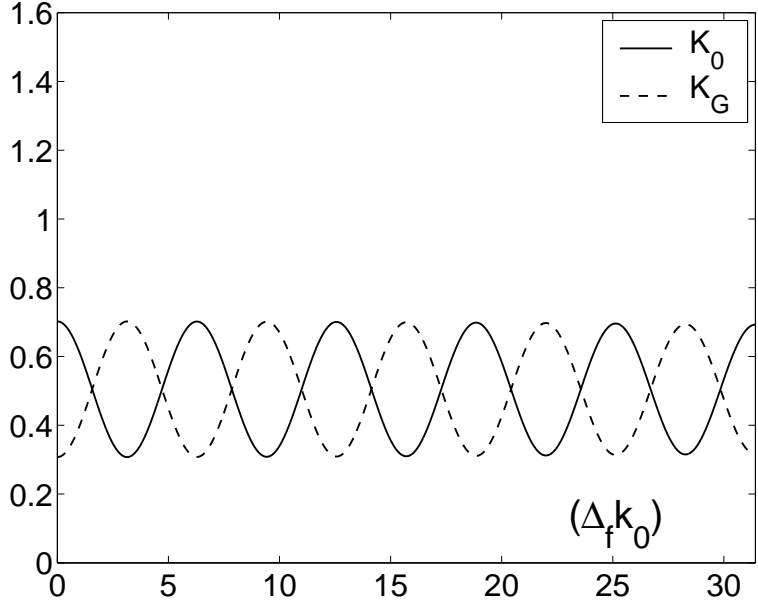}
\end{center}
\caption{\label{label}Calculated intensity oscillations of $K_{0}$
and $K_{G}$ of Eqs.(6) and (7) considering the following
parameters: $\alpha_{d}=0$, $\alpha_{f}=-(1/2)ln(28/169)=0.8988$,
$\Delta_{d}=0$ and $({\delta}k)/k_{0}=1/100$. The intensity levels
are equal due to $\alpha_{f}$. }
\end{figure}
\section{Visibilities of single and double loop interferometers}
The visibility is defined as
\begin{equation}
V=\frac{I_{max}-I_{min}}{I_{max}+I_{min}}\,\,\,\,\,.
\end{equation}
We distinguish two kinds of absorption processes: stochastic (sto)
and deterministic (det) absorption [6]. In an interferometer
stochastic absorption is realized by inserting an absorbing
material in one arm of a beam (s. Fig.(1)). Deterministic
absorption can be achieved by a chopper which blocks the beam path
periodically or by partial reduction of the beam cross section. In
the following we investigate the visibility in a single and double
loop interferometer at low interference order (coherence function
is approximately $1$).
\\
\\
a) {\it Single loop interferometer (index $1$)}: If one phase
shifter $\Delta_{d}$ and one absorber $\alpha_{d}$ is put in beam
path $(d)$, the intensity $I_{0}$ in the forward direction can be
written as
\begin{equation}
I_{0}\,{\propto}\,|\,1+e^{ik_{0}\Delta_{d}-\alpha_{d}}|^{2}=1+T_{d}+2\sqrt{T_{d}}{\,}cos(\Delta_{d}k_{0})\,\,\,\,\,,
\end{equation}
where $T_{d}=e^{-2\alpha_{d}}$ is the transmission probability of
beam $(d)$ and $\Delta_{d}k_{0}$ is the phase difference between
the two beam paths. The visibility is therefore
\begin{equation}
V_{sto1}=\frac{2\sqrt{T_{d}}}{1+T_{d}}\,\,\,\,\,.
\end{equation}
For deterministic absorption we get
\begin{equation}
I_{0}\,{\propto}\,|\,1+e^{ik_{0}\Delta_{d}}|^{2}T_{d}+(\,1-T_{d})=1+T_{d}+2\,T_{d}{\,}cos(\Delta_{d}k_{0})\,\,\,\,\,\,\,\,,
\end{equation}
\begin{equation}
V_{det1}=\frac{2\,{T_{d}}}{1+T_{d}}\,\,\,\,\,.
\end{equation}
The difference between stochastic and deterministic absorption in
a single loop neutron interferometer is drawn in Fig.(4) and has
been discussed in [6].
\\
\\
b) {\it Double loop interferometer (index $2$)}: If $\Delta_{d}$
and $\alpha_{d}$ is placed in path $(d)$ and no phase shifter or
absorber in beam path $(f)$ (see Fig.(1)), the intensity is given
as (see Eq.(6)):
\begin{equation}
K_{0}\,{\propto}\,|\,2+e^{ik_{0}\Delta_{d}-\alpha_{d}}|^{2}=T_{d}+4\,[\,1+\sqrt{T_{d}}{\,}cos(\Delta_{d}k_{0})]\,\,\,\,\,,
\end{equation}
\begin{equation}
V_{sto2}=\frac{4\sqrt{T_{d}}}{4+T_{d}}\,<\,1\,\,\,\,.
\end{equation}
The deterministic case may be expressed as
\begin{equation}
K_{0}\,{\propto}\,|\,2+e^{ik_{0}\Delta_{d}}|^{2}T_{d}+4\,(\,1-T_{d})=T_{d}+4\,[\,1+{T_{d}}{\,}cos(\Delta_{d}k_{0})]\,\,\,\,\,,
\end{equation}
\begin{equation}
V_{det2}=\frac{4\,{T_{d}}}{4+T_{d}}\,<\,1\,\,\,\,.
\end{equation}
In general, $V_{sto1}>V_{sto2}$ and $V_{det1}>V_{det2}$ for
$0{\le}T_{d}{\le}1$ (Fig.(4)).
\begin{figure}
\begin{center}
\includegraphics[angle=0,width=10cm]{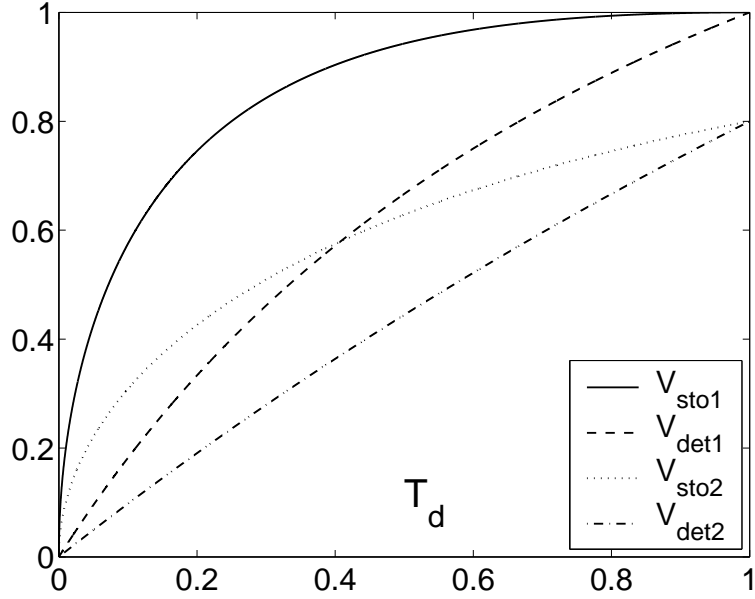}
\end{center}
\caption{\label{label}Visibilities of single and double loop
interferometers. The visibilities $V_{sto1}$ for stochastic
(Eq.(11)) and $V_{det1}$ for deterministic (Eq.(13)) absorption in
a single loop interferometer are shown as a function of
transmission probability $T_{d}$ in beam path $(d)$. In the double
loop interferometer no phase shifter $\Delta_{f}$ and no absorber
$\alpha_{f}$ is in beam path $(f)$. In this case the visibilities
$V_{sto2}$ and $V_{det2}$ of Eqs.(15) and (17) are smaller than in
a single loop device.}
\end{figure}
An interesting case arises, if a phase shifter $\Delta_{f}$ (and
no absorption $\alpha_{f}$) is inserted in beam path $(f)$. The
intensity then becomes
\begin{eqnarray}
K_{0}\,{\propto}\,|\,e^{ik_{0}\Delta_{d}-\alpha_{d}}+1+e^{ik_{0}\Delta_{f}}|^{2}=\nonumber\\=\,T_{d}+2\,\sqrt{T_{d}}{\,}[cos(\Delta_{d}k_{0})+cos(\Delta_{d}k_{0}-\Delta_{f}k_{0})]+4\,cos^{2}(\Delta_{f}k_{0}/2)\,\,\,\,\,.
\end{eqnarray}
If $\Delta_{f}=0$, then Eq.(18) reduces to Eq.(14). If
$\Delta_{f}k_{0}=(2n+1)\pi$, then $e^{i\Delta_{f}k_{0}}=-1$ and
$K_{0}{\propto}T_{d}$. If
$2n\pi{\le}\Delta_{f}k_{0}{\le}(2n+1)\pi$, maxima of $K_{0}$ are
at $\Delta_{d}k_{0}=\Delta_{f}k_{0}/2+2n\pi$ and minima at
$\Delta_{d}k_{0}=\Delta_{f}k_{0}/2+(2n+1)\pi$. These maxima and
minima are interchanged in the range
$(2n+1)\pi{\le}\Delta_{f}k_{0}{\le}(2n+2)\pi$, and it is not
necessary to consider this case separately. One gets:
\begin{equation}
K_{0,\frac{max}{min}}{\propto}\,T_{d}{\pm}4\,cos(\Delta_{f}k_{0}/2)[\sqrt{T_{d}}{\pm}cos(\Delta_{f}k_{0}/2)]\,\,\,\,\,,
\end{equation}
where the upper sign belongs to the maximum and the lower sign to
the minimum intensity. The visibility is
\begin{equation}
V_{sto2\Delta_{f}}=\frac{4\,\sqrt{T_{d}}{\,}cos(\Delta_{f}k_{0}/2)}{4\,cos^{2}(\Delta_{f}k_{0}/2)+T_{d}}\,\,\,\,\,.
\end{equation}
If $\Delta_{f}k_{0}=0$, then $V_{sto2\Delta_{f}}=V_{sto2}$. If
$\Delta_{f}k_{0}=2\pi/3$, then $V_{sto2\Delta_{f}}=V_{sto1}$. In
particular we emphasize the fact that the maximum of the function
$V_{sto2\Delta_{f}}$ is at $T_{d}=4\,cos^{2}(\Delta_{f}k_{0}/2)$
where $V_{sto2\Delta_{f}}=1$. For a given transmission $T_{d}$ in
beam path $(d)$ a phase shift
$\Delta_{f}k_{0}=2\,arccos(\sqrt{T_{d}}/2)$ in path $(f)$ should
be chosen in order to achieve a visibility of $1$ (Fig.(5) and
Fig.(6))! Setting $T_{d}=1$ the relation $V_{sto2\Delta_{f}}$ of
Eq.(20) reduces to a formula which already has been discussed in
[3] (see Fig.(6)).
\\
\\
An analogous equation can be derived for the deterministic case
which reads (Fig.(7))
\begin{equation}
V_{det2\Delta_{f}}=\frac{4\,{T_{d}}{\,}cos(\Delta_{f}k_{0}/2)}{4\,cos^{2}(\Delta_{f}k_{0}/2)+T_{d}}\,<\,1\,\,\,\,\,\,\,,\,\,for\,\,\,\,
T_{d}<1\,\,\, .
\end{equation}
In case of deterministic visibility the value of $1$ cannot be
achieved for $T_{d}<1$.
\\
\\
In order to compare intensities and to assess possibilities for
measurements, Figs.(8a,b,c) present values of $K_{0}$ and $K_{G}$
as a function of the phase shift $\Delta_{d}k_{0}$ in beam path
$(d)$. From these figures it can be concluded that in spite of
strong absorption small signals can be measured with visibility
$1$ and with sufficient intensity by adequately adjusting the
phase shifter in the second loop of the interferometer. However,
as shown in the next section, the described procedure specifies an
ideal situation and there is a limit for measuring weak signals
using such a technique.
\begin{figure}
\begin{center}
\includegraphics[angle=0,width=10cm]{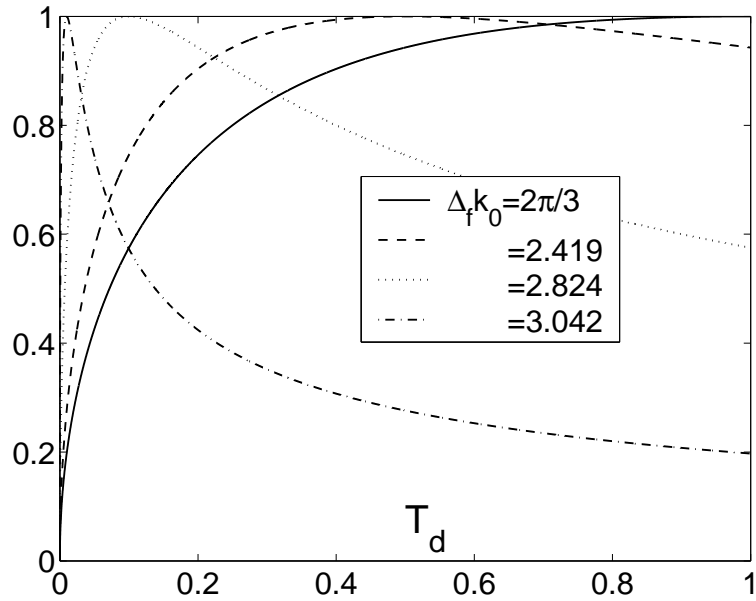}
\end{center}
\caption{\label{label}Visibility $V_{sto2\Delta_{f}}$ (Eq.(20)) in
a double loop interferometer as a function of transmission
probability $T_{d}$ in beam path $(d)$. A phase shift
$\Delta_{f}k_{0}$ is applied in beam path $(f)$. The visibility
attains the value $1$ for $T_{d}=4\,cos^{2}(\Delta_{f}k_{0}/2)$.}
\end{figure}
\begin{figure}
\begin{center}
\includegraphics[angle=0,width=10cm]{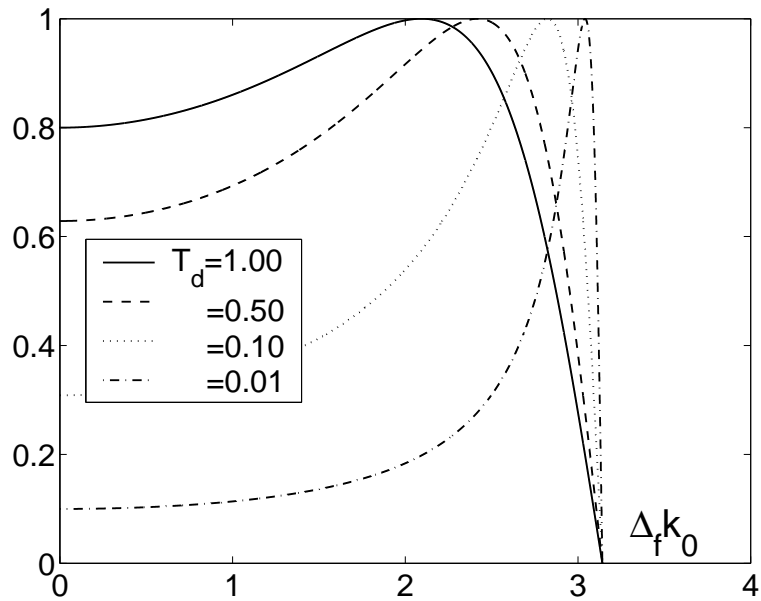}
\end{center}
\caption{\label{label}Visibility $V_{sto2\Delta_{f}}$ (Eq.(20)) in
a double loop interferometer as a function of phase shift
$\Delta_{f}k_{0}$ in beam path $(f)$. A transmission probability
$T_{d}$ is applied in beam path $(d)$. The visibility attains the
value $1$ for $T_{d}=4\,cos^{2}(\Delta_{f}k_{0}/2)$, where the $4$
parameters $T_{d}$ correspond to those $4$ parameters
$\Delta_{f}k_{0}$ in Fig.(5) via this equation.}
\end{figure}
\begin{figure}
\begin{center}
\includegraphics[angle=0,width=10cm]{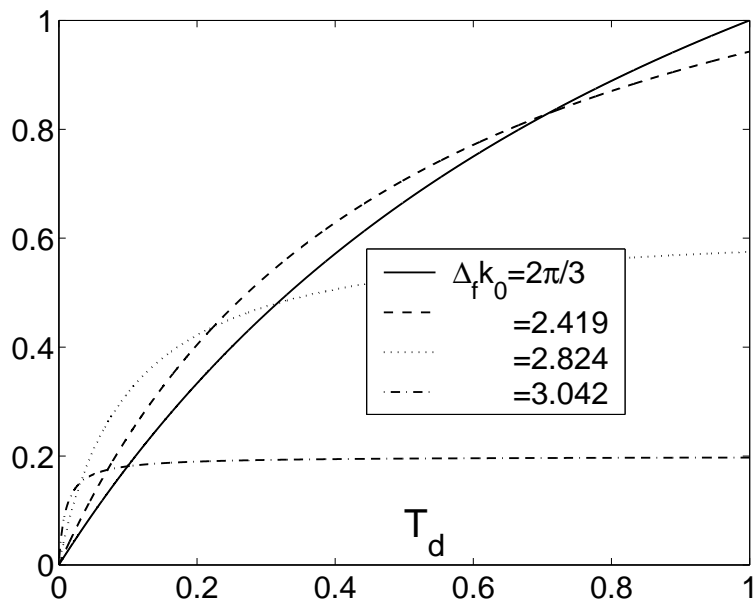}
\end{center}
\caption{\label{label}Visibility $V_{det2\Delta_{f}}$ (Eq.(21)) in
a double loop interferometer as a function of transmission
probability $T_{d}$ in beam path $(d)$. A phase shift
$\Delta_{f}k_{0}$ is applied in beam path $(f)$. Contrary to
$V_{sto2\Delta_{f}}$ this function $V_{det2\Delta_{f}}$ cannot
attain a value of $1$ for $T_{d}<1$.}
\end{figure}
\begin{figure}
\begin{center}
\includegraphics[angle=0,width=10cm]{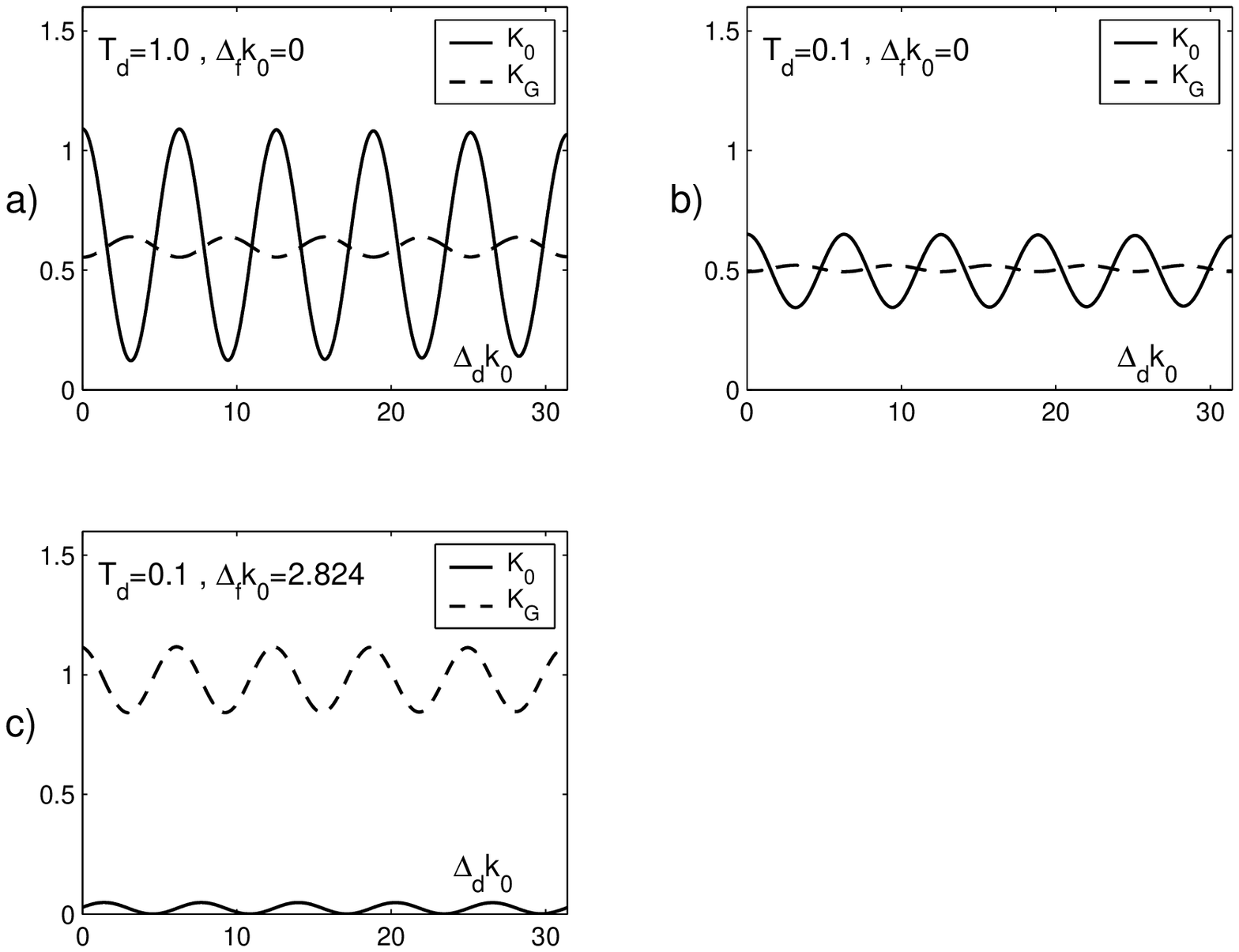}
\end{center}
\caption{\label{label}Intensities $K_{0}$ and $K_{G}$ behind the
double loop interferometer (see Fig.(1)) as a function of the
phase shift $\Delta_{d}k_{0}$ in beam path $(d)$. a) No absorption
in beam $(d)$ ($T_{d}=1$) and no phase shift in beam $(f)$
($\Delta_{f}k_{0}=0$). Fig.(8a) can be compared to Fig.(2). Note
that in Fig.(8a) the intensities $K_{0}$ and $K_{G}$ are plotted
as a function of $\Delta_{d}k_{0}$. b) Absorption in beam $(d)$
(transmission $T_{d}=0.1$) and no phase shift in beam $(f)$. The
intensities have been attenuated accordingly. c)
$\Delta_{f}k_{0}=2.824$ has been chosen. This value corresponds to
$T_{d}=0.1$ (see Figs.(5) and (6)). $K_{0}$ shows a visibility of
$1$ because of $I_{min}=0$.}
\end{figure}
\section{Discussion and conclusion}
In this paper some intensity and visibility aspects of a double
loop neutron interferometer have been considered. In the first
part the wave functions of a triple wave function superposition in
the forward and diffracted direction behind the interferometer has
been described using the theory of dynamical diffraction in single
crystals. Phase shifters and absorbers have been taken into
account. In general the intensity levels of the forward and
diffracted beams are different. Under certain circumstances these
levels can be made equal using various absorbers. This is an
interesting new aspect because in this case the device can be
regarded as a $50/50$ beam splitter. This feature is not possible
in a single loop interferometer.
\\
\\
In the second part the visibility has been investigated. As an
another important new result it has been shown that a visibility
value of $1$ may be reached when the transmission $T_{d}$ (for
stochastic absorption processes) in the first loop is adjusted
accordingly to the phase shift $\Delta_{f}k_{0}$ in the second
loop:
\begin{equation}
T_{d}=4cos^{2}(\Delta_{f}k_{0}/2)\,\,\,\,\,\,\,{\rightarrow}\,\,\,\,V_{sto2\Delta_{f}}=1.
\end{equation}
$T_{d}=e^{-2\alpha_{d}}$ is the transmission probability in beam
$(d)$ ($\alpha_{d}$ is the absorption coefficient) and
$\Delta_{f}$ is the phase shift in beam $(f)$, where - according
to Fig.(1) - the intensity $K_{0}$ in the forward direction behind
the double loop interferometer oscillates as a function of the
phase shift $\Delta_{d}$ in beam $(d)$. Using the described method
in principle every signal in path $(d)$ can be transformed to a
signal $K_{0}$ with visibility $1$ and therefore the method could
be most interesting for small signals.
\\
\\
It should be noted that to describe real experimental situations
an additive term $I_{incoh}$ must be introduced in Eq.(18) to
represent the incoherent part of the intensity (background
intensity). The main reason for the incoherent effects is non -
interference because of crystal imperfection (not absolute
parallelism of the crystal lattice planes throughout the
interferometer) as well as lattice vibrations and small
temperature gradients. The term $I_{incoh}$ leads to a new
visibility $V^{'}_{sto2\Delta_{f}}$:
\begin{equation}
V^{'}_{sto2\Delta_{f}}=\frac{4\,\sqrt{T_{d}}{\,}cos(\Delta_{f}k_{0}/2)}{4\,cos^{2}(\Delta_{f}k_{0}/2)+T_{d}+I_{incoh}}\,\,\,\,\,.
\end{equation}
This expression is always less than $1$ for $I_{incoh}>0$ (see
also Eq.(22)). In order to measure weak signals (see Fig.(8c)) the
background intensity has to be as small as possible. This could be
a serious constraint considering real experimental conditions
using neutrons. It should be mentioned that the background
intensity has no influence on Eq.(8) concerning the $50/50$ beam
splitter system. Note also that an additional empty phase which is
a signature of an interferometer does not affect the
aforementioned considerations about the intensity and visibility.
\\
\\
It is a common problem in signal processing to measure weak
signals. The absorber serves only to generate these signals. In
spite of the constraints mentioned above for almost ideal
experimental situations Eq.(22) predicts a visibility close to
$1$. This could also be the case in systems of Mach - Zehnder
interferometers using laser beams. Here the number of counts
registered in a detector in a certain time interval obeys a
Poisson distribution with variance $(\Delta{N})^{2}=\bar{N}$,
where $\bar{N}$ is the mean number of counts. The Poisson
distribution is a signature of coherent state behavior. For small
count rates the described method may be used to measure small
signals with visibility $1$.
\\
\\
The system can easily be adapted (loop $B$ in Fig.(1)) to an
eight-port interferometer as a high sensitive homodyne detection
setup [4,7,8] as shown in Fig.(9). In this case two vacuum inputs
${\,}|0>{\,}$ have to be added and the strong local oscillator
limit can be applied to the idler beam $(a)$ when the signal beam
$(e)$ is relatively strongly attenuated by an absorber
$\alpha_{e}$. Homodyne detection in laser interferometry is a
method, in which the field amplitudes (the quadrature components)
are measured instead of the quantized intensity. The balanced
version of homodyne detection has a great practical advantage of
cancelling technical noise and the classical instabilities of the
reference field. Thereby the signal interferes with a coherent
laser beam at a well-balanced $50/50$ beam splitter. It provides
the phase reference for the quadrature measurement. The intensity
difference is the quantity of interest because it contains the
interference term of the local oscillator and the signal.
\begin{figure}
\begin{center}
\includegraphics[angle=-90,width=12cm]{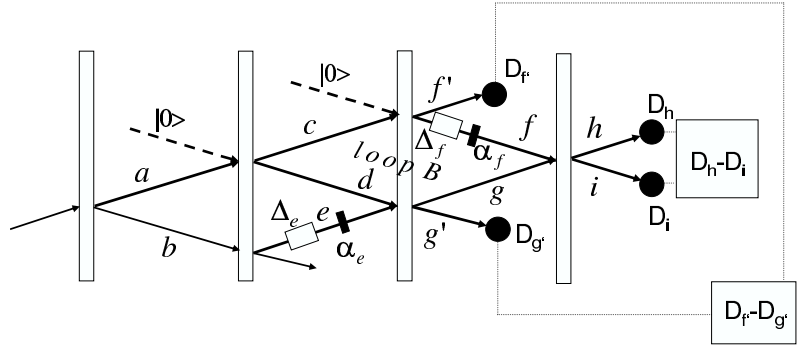}
\end{center}
\caption{\label{label}An eight-port interferometer is built up
from loop $B$ in Fig.(1). Neutron count rate differences can be
measured by the detectors $D_{h}$ and $D_{i}$ as well as $D_{f'}$
and $D_{g'}$. The phase of the strongly attenuated signal beam (e)
can be modified by a phase shift $\Delta_{e}$. The strong local
oscillator is given by the idler beam (a).}
\end{figure}
\\
\\
Finally, as already mentioned above, we would like to point out
again that there are good reasons to assume that a similar result
can be attained in double Mach-Zehnder interferometry where
photonic beams in fiber glass are used. As a visibility near to
$1$ is of great advantage for measuring largely attenuated beams,
our approach outlines a general method to investigate weak signals
in double loop interferometric devices.
\section*{Acknowledgements}
This work was supported by the Austrian Fonds zur F\"{o}rderung
der wissenschaftlichen Forschung, Wien, project SFB F-1513.

\Bibliography{<100>}

\item Rauch H and Werner S A 2000 {\it Neutron Interferometry} (Oxford: Clarendon Press)

\item Heinrich M, Petrascheck D and Rauch H 1988 {\it Z. Phys. B -
Condensed Matter} {\bf 72} 357

\item Zawisky M, Baron M and Loidl R 2002 {\it Phys. Rev. A} {\bf
66} 063608

\item Leonhardt U 1997 {\it Measuring the Quantum State of Light} (Cambridge University Press)

\item Rauch H and Petrascheck D 1978 {\it Topics in Current
Physics} {\bf 6} 303

\item Summhammer J, Rauch H and Tuppinger D 1987 {\it Phys. Rev.
A} {\bf 36} 4447

\item Freyberger M, Heni M and Schleich W P 1995 {\it Quantum Semiclass. Opt.} {\bf 7}
187

\item Walker N G and Carrol J E 1984 {\it Electron. Lett.} {\bf 20}
981

\endbib

\end{document}